%% file: main.tex
\documentclass[9pt, conference]{IEEEtran}
\usepackage[table,xcdraw]{xcolor}
\usepackage{tikz}
\usepackage{graphicx}
\usepackage{pifont}
\usepackage{multirow}
\usepackage{tablefootnote}
\usepackage{enumitem}
\usepackage{caption}
\usepackage{cite}
\usepackage{url}
\usepackage[normalem]{ulem}

\author{
\IEEEauthorblockN{
Giuseppe M. Sarda \IEEEauthorrefmark{1}\IEEEauthorrefmark{2}, 
Nimish Saha \IEEEauthorrefmark{1}, 
Abubakr Nada \IEEEauthorrefmark{2}, 
Debjyoti Bhattacharjee \IEEEauthorrefmark{2}, 
Marian Verhelst \IEEEauthorrefmark{1}\IEEEauthorrefmark{2}}
\IEEEauthorblockA{
\IEEEauthorrefmark{1}KU Leuven, Leuven, Belgium
\IEEEauthorrefmark{2}imec, Leuven, Belgium \\
Email: giuseppe.sarda@esat.kuleuven.be}
}

\newcommand*\circled[2]{\tikz[baseline=(char.base)]{
            \node[shape=circle,draw,inner sep=1pt, fill=#1] (char) {#2};}}
            
\definecolor{hwl}{RGB}{255,128,128}
\definecolor{lps}{RGB}{135,222,135}
\definecolor{dms}{RGB}{128,179,255}
\definecolor{int}{RGB}{230,230,230}

\begin{document}

\everypar{\looseness=-1}
\linepenalty=1000
\linespread{0.9}

\pagenumbering{arabic}
\pagestyle{plain}
\title{Decoupled Control Flow and Data Access in RISC-V GPGPUs}

\maketitle

\input{text/0_abstract}

\IEEEpeerreviewmaketitle

\input{text/1_intro}
\input{text/2_background}
\input{text/3_analysis}
\input{text/4_implementation}
\input{text/5_programming_model}

\input{text/6_evaluation}

\input{text/7_conclusion}

\bibliographystyle{unsrt}
\bibliography{sample-base}
\end{document}

%% file: text/0_abstract.tex
\begin{abstract}
Vortex, a newly proposed open-source GPGPU platform based on the RISC-V ISA, offers a valid alternative for GPGPU research over the broadly-used modeling platforms based on commercial GPUs.
Similarly to the push originating from the RISC-V movement for CPUs, Vortex can enable a myriad of fresh research directions for GPUs.
However, as a young hardware platform, it currently lacks the performance competitiveness of commercial GPUs, which is crucial for widespread adoption.
State-of-the-art GPUs, in fact, rely on complex architectural features, still unavailable in Vortex, to hide the micro-code overheads linked to control flow (CF) management and memory orchestration for data access.
In particular, these components account for the majority of the dynamic instruction count in regular, memory-intensive kernels, such as linear algebra routines, which form the basis of many applications, including Machine Learning. 
To address these challenges with simple yet powerful micro-architecture modifications,
this paper introduces decoupled CF and data access through 1.) a hardware CF manager to accelerate branching and predication in regular loop execution and 2.) decoupled memory streaming lanes to further hide memory latency with useful computation.
The evaluation results for different kernels show 8$\times$ faster execution, 10$\times$ reduction in dynamic instruction count, and overall performance improvement from 0.35 to 1.63 $\mathrm{GFLOP/s/mm^2}$. Thanks to these enhancements, Vortex can become an ideal playground to enable GPGPU research for the next generation of Machine Learning.
\end{abstract}

%% file: text/1_intro.tex
\section{Introduction}

In the last decade, GPGPUs have become a fundamental component in many computing systems, transitioning from a dedicated unit for graphics acceleration to a general-purpose computing workhorse for highly parallel tasks.
High-level GPU modeling simulation frameworks have been used in research to greatly 
push GPGPU's performance, efficiency, and scalability in many computing contexts~\cite{gpgpusim, accelsim, nvarchsim, gem5-gpu2}.
The GPGPU-Sim~\cite{gpgpusim} and Accel-Sim~\cite{accelsim} simulation frameworks, mostly validated on NVIDIA architectures, played a key role in this process: they model GPU architectures with more and more accuracy, allowing for a thorough analysis of optimization possibilities. \cite{dwf, mta, cta, dac,darsie} are only a few of the many papers in the GPU community using GPGPU-Sim.
Although such models of commercial GPUs represent the state-of-the-art, they lack full-scale RTL validation, limiting the methodology to high-level modeling in the academic community.

\begin{figure}
    \centering
    \includegraphics[width=0.9\linewidth]{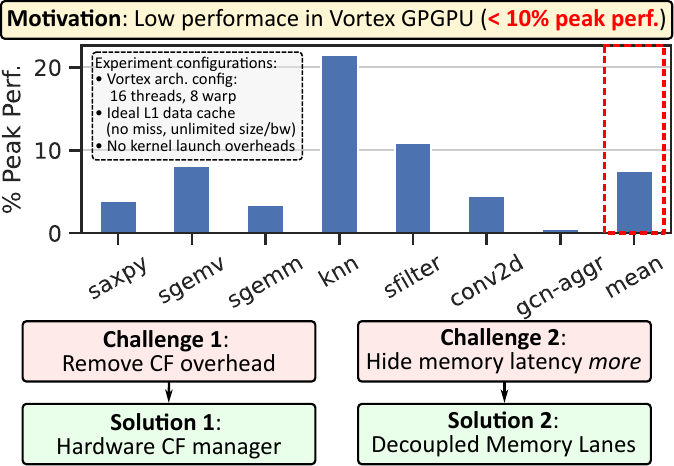}
    \caption{
    Vortex performance on various kernels, normalized to peak performance (top); roots for this performance degradation coupled to the proposed innovations targeted in this paper (bottom).
    }
    \label{fig:motivation}
\end{figure}

The Vortex GPGPU \cite{vortex,skybox}, based on the RISC-V ISA, was proposed as an alternative platform for GPGPU architecture research.
Vortex provides a full open-source GPGPU stack, from RTL implementation all the way to the compiler and software library.
Similar to the push originating from the RISC-V movement for CPUs, Vortex can create the ideal playing ground for new architecture innovations, at least if its performance is in line with the modern GPU. 
However, today, this open-source GPGPU core lacks a thorough assessment of its performance. 
Fig.~\ref{fig:motivation} shows the performance that a Vortex instantiation with 16 threads and 8 warps per core achieves on various kernels, normalized to its peak performance.
Out-of-the-box, we found that Vortex only achieves 10\% of its peak performance (i.e., 1 FLOP/cycle per thread). 
In this paper, a deeper assessment of the execution traces will further identify the loop control flow (CF) management overhead (both for execution CF and predication) and memory orchestration as the main causes of performance degradation  (see Section III).
\looseness=-1

Taking a look at commercial GPUs~\cite{a100,h100}, we observed that overhead instructions are still present, yet they are effectively hidden through many warps running in parallel (e.g. 64 per NVIDIA Streaming Multiprocessors) and through superscalar wide-issue widths. However, from the architectural point of view, superscalar features are complex, area-expensive, and only effective under a high degree of instruction level parallelism (ILP)~\cite{cgo}. 

\textbf{To eliminate overhead instructions} related to CF and memory orchestration, we implement decoupled CF and data access by extending Vortex with light-weight and simple, yet powerful micro-architecture enhancements consisting of:
1.) a hardware CF manager consisting of a zero-overhead loop control (ZOLC) block 
that drives execution of regular loops eliminating the need for control flow instructions and a loop predication stack (LPS) that applies fine-grain control over active threads through masking; and
2.) decoupled memory streaming lanes (DMSL) that perform non-speculative operands prefetch in-local buffers at the issue stage. 

We validate innovations on different kernels to assess the achievable performance gains, scalability, and demonstrate that the proposed techniques do not come at the expense of versatility, but benefit a wide range of workloads.
On average, we measure 
a 
8$\times$ speedup, 10$\times$ reduction in dynamic instruction count, and an improvement to 50\% pipeline back end utilization. From our results we show that, to match the performance of 1 of our extensions-enhanced cores, 8 baseline Vortex cores would be needed, demonstrating the effectiveness of the proposed innovations.

%% file: text/2_background.tex
\section{Background - The Vortex GPGPU}
To better understand this work, the following section gives an overview of the Vortex architecture, comparing it with equivalent features in NVIDIA GPUs.
The Vortex GPGPU is a fully open-source platform that extends the RISC-V ISA~\cite{riscv} to support the SIMT execution model. 
Vortex includes full RTL architecture implementation, a cycle-accurate functional model, the software stack (runtime kernel library and OpenCL implementation based on POCL\cite{pocl}), and an LLVM-based \cite{llvm} compiler.
The hardware architecture is hierarchically divided into clusters sharing the same L3 cache, and cores sharing the same L2 cache.
A Vortex core corresponds to an NVIDIA SMs with 1 subcore, characterized by a scratchpad shared memory, L1 data and instruction cache. A cluster maps to a collection of SMs sharing L2. The number of clusters, cores, warps, and threads is configurable at design time in Vortex, allowing for different GPU platform size and performance.
\looseness=-1

%% file: text/3_analysis.tex
\section{Vortex code analysis and related works}
\label{sec:code}

\begin{figure}
    \centering
    \includegraphics[width=0.9\linewidth]{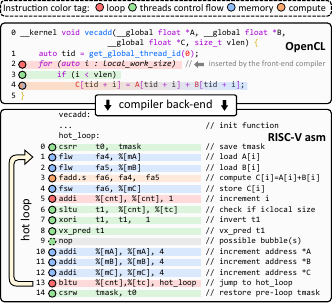}
    \caption{The OpenCL \texttt{vecadd} code compiled in RISC-V assembly.}
    \label{fig:code}
\end{figure}

In this section, we analyze in depth the bottlenecks in the current Vortex implementation and review the SotA techniques to tackle them, discussing their applicability to the Vortex platform. 
Fig. \ref{fig:code} provides a representative kernel to assess the sources of performance degradation in Vortex microcode; 
To support the GPGPU workload organization mode, the POCL compiler \cite{pocl} transforms the OpenCL code body (top) by adding {\em global} and {\em local} thread groups (grids and blocks in NVIDIA terminology). 
Thread groups (i.e. portions of the total work to be executed) are then equally distributed across warps in the hardware architecture, and during the execution, each warp iterates over the assigned groups. 
The loop routine is highlighted in the RISC-V assembly in Fig. \ref{fig:code}; the assembly instructions are color-coded into 4 different groups: 
1.)~\tikz\draw[black,fill=hwl] (0,0) circle (.5ex);~loop instructions increment the iteration count and jump back if the loop boundary condition is not met, 
2.) \tikz\draw[black,fill=lps] (0,0) circle (.5ex); predication instructions evaluate and update the active threads within a warp, 
3.) \tikz\draw[black,fill=dms] (0,0) circle (.5ex); memory code loads and stores data in the register file/cache, and takes care of updating memory pointers, and 
4.) \tikz\draw[black,fill=orange] (0,0) circle (.5ex); the compute instruction(s) evaluates the result.

While functional to the execution, red \tikz\draw[black,fill=hwl] (0,0) circle (.5ex); and green \tikz\draw[black,fill=lps] (0,0) circle (.5ex); instructions introduce overhead; 
blue instructions \tikz\draw[black,fill=dms] (0,0) circle (.5ex); load data right before compute: 
on one hand coarse-grained multi-threading (i.e. warps) in GPUs helps hiding memory latency and avoids data dependency stalls, 
but, on the other hand, 
warps can only hide latency up to a certain level 
and increasing support for more warps is area expensive \cite{nomorenoless}.
To improve performance, we identify each bottleneck and review how existing attempts to address these.

\tikz\draw[black,fill=hwl] (0,0) circle (.5ex);~\textbf{Loop CF overhead:} CF in loops relies on an iteration counter and branch instructions that check whether the loop boundary condition is met (fall-through) or the execution needs to jump back and iterate again.
To reduce the CF instruction overhead, loop unrolling executes multiple iterations within one loop body iteration \cite{loop-unrol}.
Another approach is 
zero-overhead loop control (ZOLC) or hardware loops, which updates the iteration counter and detects the fall-through condition in hardware. 
Such hardware loops are a common enhancement in DSP processors \cite{DSP,DSP2,DSP3} and CPUs \cite{ZOLC1, ZOLC2, pulp, ARM, ohw, snitch}.
In GPUs, hardware loops were not attractive as kernel iterations are distributed across highly parallel hardware resources, and in-kernel loops can represent an overhead depending on the launch-time arguments (e.g. if the iteration count is too small). Yet, in Vortex, work group execution is folded in a (regular) loop execution (see line \texttt{2} of OpenCL code in Fig.~\ref{fig:code}), resulting in significant CF management overhead, which appears in all kernels execution.

\tikz\draw[black,fill=lps] (0,0) circle (.5ex);~\textbf{Predication overhead:} 
Predication is a fundamental aspect of single-instruction multiple-data (SIMD) architectures; masking ensures correct and fine control flow across execution lanes. 
During loop execution, the iterations ``issued" by the shared front end of the pipeline is the same for all threads. In case one thread meets the end condition earlier than the others, the program must take care of masking through the following steps (linked to instructions from Fig.~\ref{fig:code}): a.) save the initial thread mask in a temporary register (line \texttt{0}), b.) evaluate for each iteration which threads are still active (lines \texttt{6} and \texttt{7}), c.) update the thread mask configuration register (line \texttt{8}) and wait until the update is effective before proceeding to avoid read-after-write (RAW) hazards on the thread mask (line \texttt{9}), d.) and, at the end of the loop, restore the initial thread mask value (line \texttt{14}).
This procedure represents a significant overhead, especially in the case of multiple nested loops (e.g., in matrix multiplication or convolutions) and deep pipelines.
DAC \cite{dac} introduced predicate expansion units that allow a scalar increment of the loop iterator: 
A leader warp executes the iterator comparison with the loop boundary conditions, then the result is ``expanded'' in a bit-mask through a dedicated hardware block. Unfortunately, execution highly depends on the scheduling of this leader warp, making this technique prone to stalls.

\begin{figure*}
    \centering
    \includegraphics[width=0.95\linewidth]{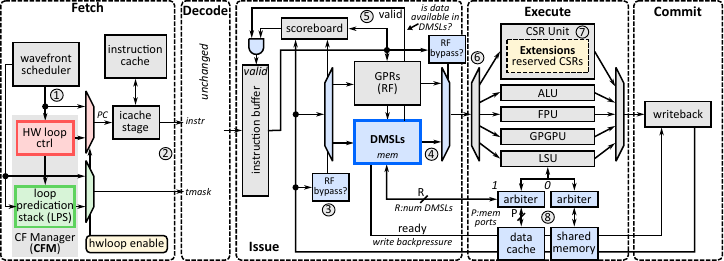}
    \caption{System level integration and extension units interaction with the 
    GPGPU pipeline. The CF Manager (CFM) is placed in the fetch stage, while Decoupled Memory Streaming Lanes (DMSLs) sit in the issue stage.}
    \label{fig:integration}
    \vspace{-0.2cm}
\end{figure*}

\tikz\draw[black,fill=dms] (0,0) circle (.5ex);~\textbf{Memory orchestration overheads:} In general-purpose processors, load and store instructions move data between the register file~(RF) and the memory system.
In regular loops, every memory instruction is followed by an address update of a fixed stride.
To avoid this extra instruction, post-increment load and store instructions, present in several processors~\cite{HP,arm7,tensilica,pulp}, perform the memory operation and the address update at the same time. 
Similarly, NVIDIA GPUs support vector load and store~\cite{nvidia-vec} and explicitly programmed memory transfers (e.g. TMA~\cite{tma}) to minimize RF/memory iterations as well as address calculation overhead~\cite{nvidia-vec2,nvidia-vec3}. A similar approach was proposed by \cite{abubakr}, but tailored to loading and offloading data from the tensor core for only matrix multiplication execution.

Another issue is that load instructions are executed \textit{too close} to compute. Long memory latency can, in fact, cause data dependency stalls even in GPUs. As a solution, \cite{mta, prefetch1, prefetch2, prefetch3, cta} proposed prefetching techniques that exploit GPU strided access regularity and interleaved warp execution to achieve high-accuracy and timely prefetch requests.
While speculative prefetching is effective to reduce memory latency, it 
can increase energy consumption due to wrong predictions~\cite{gdac}, plus it does not hide last-level data cache bank contention.
Arnau et al. \cite{gdac}, based on \cite{dac82} and also used in 
\cite{dva,outrider,nsp}, proposed decoupled access-execute with non-speculative prefetches for GPUs. Yet, their technique only targets game rendering in mobile GPUs and is limited to in-cache prefetching, missing the chance of hiding L1 bank contention as well. 
Some CPU-based systems~\cite{snitch,manticore,occamy}, use data streamers for nonspeculative, in-local buffer prefetching~\cite{ssr}. 
Nevertheless, such streamers are designed for single-threaded scalar execution and are tightly coupled to simple scratchpad memories~(TCM). This approach cannot be directly applied to complex cache designs for SIMT architectures, such as Vortex.
\looseness=-1

%% file: text/4_implementation.tex
\section{Proposed Vortex enhancements}

To remove loop CF overhead and allow decoupled access/execute in the Vortex GPGPU, this work introduces, respectively, a hardware Control Flow Manager (CFM) for loops and decoupled memory streaming lanes (DMSLs).
The main idea is to configure the hardware extensions through software \textit{once} before the beginning of the hot loop(s), to avoid executing the overhead instructions at each iteration.

\subsection{System-level Overview}

Fig. \ref{fig:integration} highlights the integration of the new hardware blocks for CFM and DMSL in the Vortex GPGPU system. 
The fetch stage integrates the \textit{hardware CFM}, consisting of hardware loops, that increment the iteration counter and jump back to the loop start program counter (PC), and a special predication stack, called loop predication stack (LPS), to fine-grain control active threads in nested loops. The hardware loops and the LPS receive information from the warp scheduler and the immediate post-dominator stack \circled{int}{1}. During loop execution, they override the PC and thread mask before going to the decode stage \circled{int}{2}.\looseness=-1

In the issue stage \textit{DMSLs} are able to prefetch data from memory and update memory addresses autonomously:
when an operand is mapped to a memory stream lane \circled{int}{3}, data comes directly from the streamers, bypassing the RF \circled{int}{4}. More precisely, before issuing an instruction, the instruction buffer checks if the operands are mapped into a DMSL, and in case they are, looks up if data is available for \textit{all} the threads in the scheduled warp \circled{int}{5}.
If this condition is true, the instruction is then dispatched to the correct execution unit \circled{int}{6}; if not all data is present, the instruction is suspended and another one is evaluated for execution. If none of the warps are active, the system is stalled until the DMSLs have fetched the required data to proceed. 
This backpressure mechanism works similarly to RAW hazard prevention in the scoreboard.

Finally, in the execution stage, part of the Configuration and Status Registers (CSRs) are reserved to configure all the extensions \circled{int}{7}, according to the programming model explained in Section~\ref{sec:pmod}. The core memory system was also modified to support independent concurrent requests from the DMSLs and the Load and Store Unit~(LSU). To further improve performance and reduce request conflicts, the bandwidth of the data cache was increased to a configurable number of independent ports \circled{int}{8}; the implementation details and the performance/overhead impact of the multiport data cache are discussed in Sections \ref{sec:memsys} and \ref{sec:eval}. \looseness=-1

\subsection{Hardware Control Flow Manager (CFM)}

The hardware CFM uses a hardware loop block and a loop predication stack (LPS) to accelerate predication and branch instructions in the pipeline fetch stage. The hardware loop replaces micro-instructions tagged in red (\tikz\draw[black,fill=hwl] (0,0) circle (.5ex);) in Fig. \ref{fig:code}, while the LPS eliminates instructions tagged in green \tikz\draw[black,fill=lps] (0,0) circle (.5ex);).

\subsubsection{Hardware loops}
\begin{figure}[]
    \centering
    \includegraphics[width=0.85\linewidth]{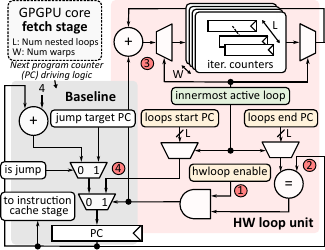}
    \caption{Hardware loops unit microarchitecture. The unit removes loop CF overhead by accelerating loop branch instructions and incrementing the loop iteration counter at the fetch stage.} 
    \label{fig:hwl}
\end{figure}

The hardware loop extension block keeps track of the state of up to \textit{L} nested loops, where \textit{L} can be configured at design time. 
Fig. \ref{fig:hwl} depicts the hardware architecture of the HW loop control block. 
After a new loop is configured and enabled \circled{hwl}{1}, the block evaluates the state of the warp selected by the scheduler, and, in case the current instruction PC matches the end of the innermost running loop \circled{hwl}{2}, the hardware updates the corresponding loop counter \circled{hwl}{3} and drives the next PC \circled{hwl}{4}. 
As the processor pipeline front end is shared across all threads, the execution iterates over the loop body until all the threads of the wave have met the end condition. Correct lane activation (predication) can be ensured in our design by explicitly programming the thread mask via software or through the dedicated LPS extension.

\subsubsection{Loop predication stack (LPS)}

\begin{figure}[]
    \centering
    \includegraphics[width=0.9\linewidth]{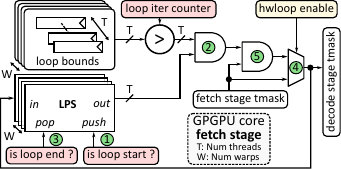}
    \caption{Loop predication stack (LPS) microarchitecture. The stack removes predication overhead by applying fine-grain control over active threads in nested loops.}
    \label{fig:lps}
\end{figure}

The LPS extension is tightly coupled to the hardware loop block and is also located in the fetch stage.
The LPS aims to save and restore thread masks across nested loops in a stack-like behavior and avoids the overhead of predication at each kernel iteration.
Note that using the LPS inherently prevents any thread mask RAW hazard from occurring; when implementing prediction via software, the predication instruction goes through the whole processor pipeline before it is effective; LPS, on the contrary, evaluates the active thread mask at the fetch stage avoiding the need for nop bubbles.
Furthermore, saving/restoring the thread mask before/after the loop execution is a minimal overhead in single-loop kernels like \texttt{vecadd}, but it becomes relevant in the case of multiple nested loops.
The implementation of the LPS is shown in Fig. \ref{fig:lps}. When the beginning of a loop is detected, the thread mask generated in the fetch stage is pushed onto the stack \circled{lps}{1}. 
Then, the reference mask on the front of the stack is updated through a bitwise AND operation with the control of the current loop under execution. The AND will disable threads that have already completed their loop iterations \circled{lps}{2}. 
When all end conditions are met, the loop is completed and the thread mask associated with that loop level is popped from the front of the stack \circled{lps}{3}. 
If no loops are still executing, then the regular fetch-stage thread mask is used to drive the execution in the rest of the pipeline~\circled{lps}{4}. \looseness=-1

\subsubsection{Control flow divergence} 
To support an {\em if-then-else} case happening within a loop (that can potentially lead to thread divergence), the loop iteration thread mask, generated by the first bitwise AND operator \circled{lps}{2}, is further compared with the fetch stage mask \circled{lps}{5}, set by the divergence post-dominator stack of the baseline Vortex architecture. 
For correct operation, the join instruction must be within the loop body, i.e. the \textit{end-loop} PC must always come \textit{after} the divergent portion of the code within the loop.
Thread divergence across loops, e.g. if a for loop is contained inside a control flow divergent block, does not require specific hardware support; the execution will be correct when the loop ends \textit{before} the join instruction.

\subsection{Decoupled Memory Streaming Lanes (DMSLs)}
The second proposed Vortex optimization aims to remove explicit memory address updates and further hide memory access latency. To this end, DMSLs, in the issue stage, independently generate non-speculative prefetch requests, store data locally in dedicated FIFO queues, and update the related memory pointer through linear computation.
FIFO queues are used as a local buffer to accumulate data in order of access and hide memory latency. 
Each DMSL, \textit{R} in total, can be assigned to a different data stream operand. In the example of Fig. \ref{fig:code}, three independent DMSLs replace instructions \texttt{1}-\texttt{10} for the source operand A, \texttt{2}-\texttt{11} for B and \texttt{4}-\texttt{12} for result C. All instructions belong to the memory tag in blue \tikz\draw[black,fill=dms] (0,0) circle (.5ex);). 

\subsubsection{Microarchitecture}
\begin{figure}
    \centering
    \includegraphics[width=0.9\linewidth]{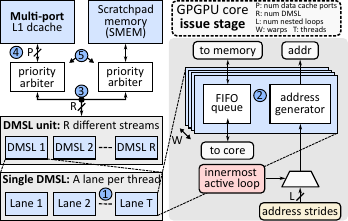}
    \caption{The DMSL unit connected to the memory system. The unit contains multiple DMSLs, each providing a streaming lane per thread. DMSLs take care of memory orchestration by non-speculative prefetching data in-local buffer and updating memory pointers.}
    \label{fig:dmsl}
\end{figure}

Fig. \ref{fig:dmsl} gives an overview of the implementation of the DMSL unit, from the top level to the data streamer used by a single thread.
Each DMSL comprises a streaming lane (SL) per thread \circled{dms}{1}. 
The SL design is inspired by \cite{ssr}, but with added support for multi-threading. 
Each warp has, in fact, its own FIFO queue and memory pointer state \circled{dms}{2}. 
FIFO memories can hold a fixed number of data elements, called \textit{credits}, configurable at design time. DMSLs can be set to read, write, or read-write mode in the case an operand is needed for multiple iterations, e.g. in accumulation.

\subsubsection{Enhanced memory system}
\label{sec:memsys}
To further increase performance, we enhanced the memory system to allow for simultaneous access to the L1 cache and shared memory. Furthermore, we increased the number of independent cache lines accessible in one cycle through (\textit{P}) ports in the L1 data cache. \textit{P} is another parameter configurable at design time. 
In the Vortex baseline, the LSU can only send one memory request per cycle when executing load and store instructions. 
Offloading data orchestration to the DMSLs, as dedicated memory traffic acceleration units, removes this limitation, allowing each thread to handle more memory requests simultaneously.
When multiple DMSLs are instantiated, each one can provide an operand per cycle, by independently accessing the memory system.
In an ideal case, every operand needed for the compute operation is mapped to a data stream with its dedicated access port to memory (cache or scratchpad). 
However, in a real-case scenario, the number of data streams, mapped in the DMSLs, should not be tied by the number of memory ports: for this reason we 1.) made both the L1 data cache and shared memory concurrently accessible by the DMSL \circled{dms}{3} 2.) extended the number of \textit{P} ports in the L1 data cache memory, where each port can simultaneously access an independent cache line \circled{dms}{4}, but 3.) made the access \textit{virtual} through a priority arbiter that selects, cycle by cycle, one valid request per port from the \textit{R} different DMSLs \circled{dms}{5}.
The priority is based on the number of valid \textit{credits} present in the DMSLs' FIFO; the arbiter will grant a memory request to the DMSL in more need (the one with fewer or more data, respectively, in the cases of read or write mode).
Finally, the first data cache port is always multiplexed with the LSU to keep compatibility with the regular execution. The LSU has higher priority over DMSLs.

%% file: text/5_programming_model.tex
\section{Programming model}
\label{sec:pmod}

\begin{figure*}
    \centering
    \includegraphics[width=0.98\linewidth]{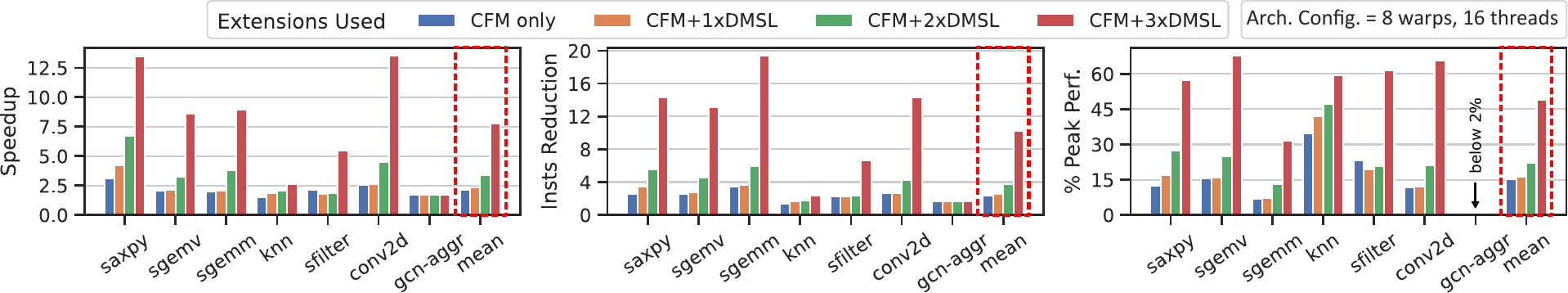}
    \caption{Speedup, dynamic instruction reduction, and achieved (\%) peak performance exploiting different extensions (bars), across different kernels (x-axis).}
    \label{fig:speedup}
\end{figure*}

The Vortex programming model has to be adapted to be able to exploit the presented extensions. The extensions are active in a specific, localized moment of loop execution, and always configured \textit{before} the beginning of the hot loop(s). This operation can be seen as moving the overhead instructions from the loop code to a one-time, low-overhead setup that happens ahead of time.
We implement these configurations through the programming of dedicated CSRs with regular, non-blocking \texttt{csrrw}, \texttt{csrrs}, and \texttt{csrrc} instructions. 
We opted for CSRs instead of ISA extensions, as CSRs enable integration without invasive modifications to the processor pipeline. 


An 8-bit address encodes the semantics for setting a unit type (e.g. CFM or DMSL), the unit ID, and the register number.
Table~\ref{tab:csrs} lists the different configurations with information regarding the unit type and programming mode. 
The CFM configurations (for each of the supported \textit{L} nested loops, are the loop start and end PC, the thread mask of the last iteration (tail tmask) to support tail execution, loop bound (e.g. the total number of iterations), the enable flag, and the loop iteration count state. 
For the DMSLs, configurations are the base memory address, the address of the mapped register in the RF (e.g. register \texttt{t1} would be \texttt{x6}), if the data is an integer or a floating point, data precision (e.g. 32, 16, 8 bit), 
and enables to start prefetching and to bypass RF accesses to DMSLs.


\begin{table}[]
\caption{Configuration registers for different units.}
\resizebox{\linewidth}{!}{ \scriptsize
\begin{tabular}{|cccccc|}
\hline
\multicolumn{3}{|c|}{\textbf{CFM}}                                                                               & \multicolumn{3}{c|}{\textbf{DMSL}}                                                              \\ \hline
\multicolumn{1}{|c|}{\textbf{Reg ID}}    & \multicolumn{1}{c|}{\textbf{Bits}} & \multicolumn{1}{c|}{\textbf{Config}} & \multicolumn{1}{c|}{\textbf{Reg ID}}    & \multicolumn{1}{c|}{\textbf{Bits}} & \textbf{Config} \\ \hline
\multicolumn{1}{|c|}{0}                  & \multicolumn{1}{c|}{31:0}          & \multicolumn{1}{c|}{start PC}        & \multicolumn{1}{c|}{0}                  & \multicolumn{1}{c|}{31:0}          & base mem. addr. \\ \hline
\multicolumn{1}{|c|}{1}                  & \multicolumn{1}{c|}{31:0}          & \multicolumn{1}{c|}{end PC}          & \multicolumn{1}{c|}{\multirow{5}{*}{1}} & \multicolumn{1}{c|}{4:0}           & RF reg.         \\ \cline{1-3} \cline{5-6} 
\multicolumn{1}{|c|}{2}                  & \multicolumn{1}{c|}{31:0}          & \multicolumn{1}{c|}{end tmask}       & \multicolumn{1}{c|}{}                   & \multicolumn{1}{c|}{6:5}           & FP/INT          \\ \cline{1-3} \cline{5-6} 
\multicolumn{1}{|c|}{\multirow{2}{*}{3}} & \multicolumn{1}{c|}{30:0}          & \multicolumn{1}{c|}{bound}           & \multicolumn{1}{c|}{}                   & \multicolumn{1}{c|}{9:7}           & precision       \\ \cline{2-3} \cline{5-6} 
\multicolumn{1}{|c|}{}                   & \multicolumn{1}{c|}{31}            & \multicolumn{1}{c|}{enable}          & \multicolumn{1}{c|}{}                   & \multicolumn{1}{c|}{10}            & prefetch en.    \\ \cline{1-3} \cline{5-6} 
\multicolumn{1}{|c|}{4}                  & \multicolumn{1}{c|}{31:0}          & \multicolumn{1}{c|}{loop state}      & \multicolumn{1}{c|}{}                   & \multicolumn{1}{c|}{11}            & redirect en.    \\ \hline
\multicolumn{1}{|c|}{5}                  & \multicolumn{1}{c|}{31:0}          & \multicolumn{4}{c|}{address stride}                                                                                                   \\ \hline
\end{tabular}}
 \vspace{-1em}
    \label{tab:csrs}
\end{table}

%% file: text/6_evaluation.tex
\section{Evaluation}
\label{sec:eval}

\subsubsection{Methodology}
We implemented the microarchitecture extensions within the event-based, cycle-accurate model (C++) of 
the Vortex platform, and also implemented the dominant hardware modifications at the RTL level. This allowed us to accurately analyze the benefits in terms of execution efficiency, as well as the expected silicon area overheads. 
Table \ref{tab:benchm} lists the benchmarks used for the evaluation. To understand how much benefit each hardware extension brings to the GPU platform, the kernels are coded to selectively activate specific hardware extension during the execution. 
When measuring the performance on Vortex, the data cache system plays an important role: the way data aligns in cache can highly impact memory bank contention and hence the execution latency. As this is strongly influenced by the size of the source/destination operands, we ran benchmarks sweeping the workload size as indicated in Table II, 
and report the average performance across each sweep. This gives more confidence towards representative 
and reliable results.

\subsubsection{Execution benefits}
Fig. \ref{fig:speedup} shows the improvements in terms of speed, instruction count and utilization of the extended Vortex compared to the baseline Vortex platform (with no penalty in simulation time), when progressively enabling more of the extensions proposed in this paper. From left to right, the results show that, with the proposed extensions, we can achieve on average 8$\times$ speedup, 10 $\times$ dynamic instruction reduction, and 50\% peak performance gains.
Looking at the results in more detail, it is possible to distinguish two trends among the benchmarks: 1.) \texttt{saxpy}, \texttt{sgemv}, \texttt{sgemm}, and \texttt{conv2d} are regular data-intensive kernels which benefit the most from the extensions, with up to 10$\times$ speedup and 15$\times$ instruction reduction. \texttt{sgemm} represents an exception, especially in the achieved peak performance, due to high bank contention at the L1 data memory. 2.) \texttt{knn} and \texttt{sfilter} are compute-intensive and have the best utilization among all the benchmarks already in the baseline version (see Fig. 1), still they reach 2.5$\times$ and 5$\times$ speedup, respectively. Graph Convoulutional Network (GCN) aggregation (\texttt{gcn\_aggr}) \cite{gcn} is a special kernel that is based on graph navigation and, hence requires indirect data access; for this reason, DMSL can't be used, but CFM alone can provide 1.7$\times$ speedup and dynamic instruction reduction. 

\subsubsection{Speedup scalability analysis}
\label{scalability}

\begin{table}[]
\caption{List of benchmarks - B:~BLAS, D:~Rodinia \cite{rodinia}, ML:~Machine Learning.}
\resizebox{\linewidth}{!}{
\begin{tabular}{cccccc}
\hline
\multicolumn{1}{|c|}{\textbf{Name}}  & \multicolumn{1}{c|}{\textbf{Type}} & \multicolumn{1}{c|}{\textbf{Parameters$^1$}}                                                     & \multicolumn{1}{c|}{\textbf{Name}}                                               & \multicolumn{1}{c|}{\textbf{Type}} & \multicolumn{1}{c|}{\textbf{Parameters$^1$}}                                                                      \\ \hline
\multicolumn{1}{|c|}{\texttt{saxpy}} & \multicolumn{1}{c|}{B}    & \multicolumn{1}{c|}{x={[}4:200:20{]}}                                               & \multicolumn{1}{c|}{\texttt{sfilter}}                                            & \multicolumn{1}{c|}{D}    & \multicolumn{1}{c|}{x*y={[}4:50:4{]}}                                                                \\ \hline
\multicolumn{1}{|c|}{\texttt{sgemv}} & \multicolumn{1}{c|}{B}    & \multicolumn{1}{c|}{\begin{tabular}[c]{@{}c@{}}x={[}4:200:20{]}\\ z=8\end{tabular}} & \multicolumn{1}{c|}{\texttt{conv2d}}                                             & \multicolumn{1}{c|}{ML}   & \multicolumn{1}{c|}{\begin{tabular}[c]{@{}c@{}}C=8 K=8 F=3x3\\ x*y={[}2:25:2{]}\end{tabular}}        \\ \hline
\multicolumn{1}{|c|}{\texttt{sgemm}} & \multicolumn{1}{c|}{B}    & \multicolumn{1}{c|}{\begin{tabular}[c]{@{}c@{}}x*y={[}4:50:4{]}\\ z=8\end{tabular}} & \multicolumn{1}{c|}{\multirow{2}{*}{\begin{tabular}[c]{@{}c@{}}\texttt{gcn\_aggr}\end{tabular}}} & \multicolumn{1}{c|}{\multirow{2}{*}{ML}} & \multicolumn{1}{c|}{\multirow{2}{*}{\begin{tabular}[c]{@{}c@{}}k=16~datasets={[}cora,\\ citeseer,pubmed{]}\end{tabular}}} \\ \cline{1-3}
\multicolumn{1}{|c|}{\texttt{knn}}   & \multicolumn{1}{c|}{D}    & \multicolumn{1}{c|}{x={[}4:200:20{]}}                                               & \multicolumn{1}{c|}{}                                                                       & \multicolumn{1}{c|}{}                    & \multicolumn{1}{c|}{}                                                                                                      \\ \hline

\multicolumn{6}{l}{\begin{tabular}[c]{@{}l@{}}$^1$Parameters indicate the sweep as {[}\texttt{start}:\texttt{end}:\texttt{step\_size}{]}, normalized\\ over the threads available in the GPGPU configuration under benchmark.\end{tabular}}                                                                                                                                              
\end{tabular}
}
\vspace{-1em}
    \label{tab:benchm}
\end{table}

\begin{figure}
    \centering
    \includegraphics[width=0.95\linewidth]{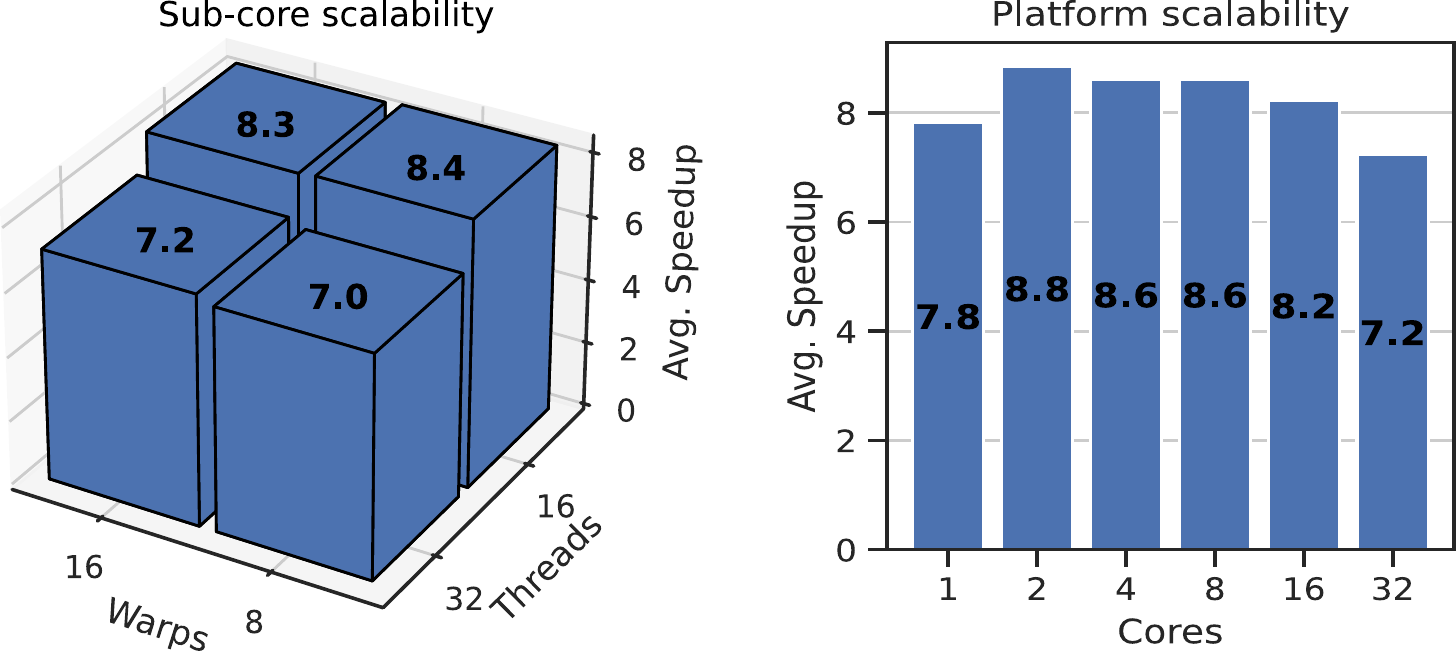}
    \caption{Scalability analysis of average speedup across Table \ref{tab:benchm} benchmarks (except \texttt{gcn\_aggr}). The left plot shows sub-core scalability in terms of warp and thread configurations, the right one shows the number of core sweep with 16 threads, 8 warps configuration (\texttt{w8t16}).\looseness=-1}
    \label{fig:scalability}
    \vspace{-0.3cm}
\end{figure}

As scalability is key in modern computing systems, we study here the \textit{speedup scalability} 
of the extended Vortex GPGPU 
when scaling up its architecture in terms of 1.) the number of threads per warp and warps per core; and 2.) the number of cores.
Fig. \ref{fig:scalability} shows the obtained results, averaged over all Table II benchmarks, and normalized to the performance achieved on the baseline Vortex platform.
The scalability results of a single core (Fig. 8 left) 
shows that the speedup compared to the baseline Vortex remains significant when scaling up the number of threads and warps. The configurations with 32 threads reach a slightly lower speedup compared to 16 threads (7$\times$ vs 8$\times$, respectively), due to an increase in the L1 data cache bank contentions. 
For the number of warps, we observe minimal variations, demonstrating that with the current extended architecture, we saturate the achievable performance.
Fig. 8~(right) demonstrates that scaling the number of cores also maintains 
the effectiveness of the proposed extensions, making them attractive in multiple applications with different performance and area targets.

\begin{figure}
    \centering
    \includegraphics[width=\linewidth]{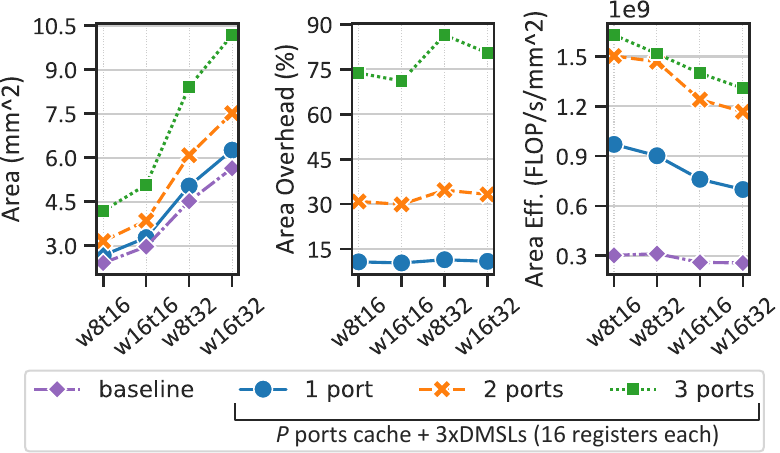}
    \caption{Core area in $\mathrm{mm^2}$ (left), area overhead in \% (center), and area efficiency in $\mathrm{FLOP/s/mm^2}$ (right) of the Vortex GPGPU with and without the proposed extensions. Plots show different warp-thread (\texttt{wXtY}) configurations. \texttt{fmadd} instruction is considered as 1 FLOP.
    }
    \vspace{-1em}
    \label{fig:area}
\end{figure}
\subsubsection{Area overhead}
\label{area_ovh}
To estimate the overhead of the extensions to the Vortex GPGPU silicon footprint, we ported the Vortex RTL to ASIC design, substituting FPGA BRAM with SRAM memories, generated with Synopsys Foundation IP memory compiler. The design was synthesized with Cadence DDI 22.35 suite in modern GlobalFoundries 22nm FDX technology at 800 MHz with different architectural configurations. 
The study focuses on the major area contributors of the proposed extensions, being 
1.) the extended bandwidth and increased bank number of the L1 data cache (still keeping the same baseline data cache memory size); and 2.) the DMSL's FIFO structures that host data locally. 

Fig. \ref{fig:area} shows the area, area overhead, and area efficiency of the proposed extension and enhanced memory system, compared to the baseline Vortex.
The system under analysis is a 4-cluster, 4-core architecture with different wavefront and thread configurations.
Area values are normalized for the total number of cores~(=16). Area efficiency is evaluated using the average performance running the benchmarks in Table \ref{tab:benchm}.
From the plots, we can observe that the local memory added by the DMSLs increases the total area of the GPGPU only by up to 15\% (1 port, center) 
, while providing more than 2$\times$ better area efficiency (right). 
Extending the data cache bandwidth, on the other hand, has a more dominant impact on the area: 1 extra port adds between 30 to 40\% more area (2 ports, center)
, but still provides $4\times$ better area efficiency. 
Finally, adding a third port 
accounts for up to 90\% area overhead, with marginal benefit on the area efficiency over the 2 ports 
solution. 
The choice of architectural configuration, in terms of thread, warp, and port count, implies an area/performance trade-off. Each knob can be adjusted according to the area budget and target performance, providing great flexibility to the designer.

\subsubsection{Comparison table} To finally assess the overall benefit of the proposed technique in terms of performance and area efficiency against the SotA, we benchmark in Table \ref{tab:comp} the Vortex platform enhanced with our hardware extensions to various instantiations of the baseline Vortex GPGPU.
Metrics for Vortex are the result of the average performance when running kernels of Table \ref{tab:benchm} (as for \ref{scalability} and \ref{area_ovh}). All Vortex architectures are based on the 8-warps, 16-threads configuration. For our enhanced system, we selected the 3-ports variant with all our extension enabled.
Numbers for Spatz \cite{spatz} and Ara2 \cite{ara2} are extrapolated from the kernels shown in the respective papers. We comment on the table in 3 points:
\\
1.) As comparison points, we first of all benchmark against the Vortex baseline architecture with core counts 1, 2, 4 and 8 
(indicated as VB \cite{vortex}). 
The results in Table~\ref{tab:comp} demonstrate that in order to match the performance of a single extensions-enhanced core, 8 
baseline Vortex cores 
would be needed, which would result in a 6.5$\times$ larger silicon footprint and 5.4$\times$ smaller area efficiency. 
\\
2.) As compiler-based loop unrolling is a common technique to reduce control flow overhead (see Section \ref{sec:code}), it is interesting to assess whether similar performance gains can be achieved through it, compared to the proposed extensions.
We therefore also benchmark against the baseline Vortex architecture with software unrolling of the lowest level loop through static Clang compiler directives \cite{loop-unrol,llvm}.
Through ablation experiments, we found 4 and 8 to be the best unrolling factors for performance, denoted as $\mathrm{VU_4}$ and $\mathrm{VU_8}$. 
The results in Table III, however, show that while loop unrolling helps for certain kernels and workload sizes, it is overall not very effective when the number of nested loops increases. Furthermore, static unrolling can't dynamically adapt to launch-time kernel parameters (e.g., in case the unrolling factor is larger than actual loop iteration count). The proposed hardware extensions, instead, are set based on launch-time parameters and better adapt to different kernel's shape; quantitatively, our work offers 4.3$\times$ better area efficiency compared to software loop unrolling.
\\
3.) We finally compare this work to Ara2 and Spatz. Our Vortex platform outperforms the application-class Ara2 core in performance and area efficiency. Spatz, instead, is a minimal sub-system that achieves comparable performances at a smaller footprint. While Vortex is a complete and versatile platform with more than 1MB of memory in three levels of cache, Spatz needs integration in a full system and packs only 128kB of L1 scratchpad memory.

\begin{table}[]
\caption{Comparison table for Performance and Area efficiency across different platforms. Results are normalized for 800MHz clock frequency. \texttt{fmadd} instruction is considered as 1 FLOP. 
}
\resizebox{\linewidth}{!}{
\begin{tabular}{|c|crrr|}
\hline
\rowcolor[HTML]{FFFFFF} 
VB \cite{vortex}                                                           & \multicolumn{4}{l|}{\cellcolor[HTML]{FFFFFF}Vortex Baseline architecture}                                                                                                                                                                                                                                                                                                                                               \\ \hline
$\mathrm{VU_F}$  \cite{loop-unrol}                                                        & \multicolumn{4}{l|}{\begin{tabular}[c]{@{}l@{}}Vortex Baseline architecture with Clang loop unrolling enabled\\ F is the unrolling factor used in the experiment\end{tabular}}                                                                                                                                                                                                                                          \\ \hline
Spatz  \cite{spatz}                                                        & \multicolumn{4}{l|}{\begin{tabular}[c]{@{}l@{}} Vector core based on Snitch \cite{snitch}. 2-cores, 8-lanes system. \end{tabular}}                                                                                                                                                                                                                                          \\ \hline
Ara2  \cite{ara2}                                                        & \multicolumn{4}{l|}{\begin{tabular}[c]{@{}l@{}} Application-class vector core (RVV 1.0). 16-lane, 512 VL system \end{tabular}}                                                                                                                                                                                                                                          \\ \hline
\rowcolor[HTML]{EFEFEF} 
\textbf{\begin{tabular}[c]{@{}c@{}}This\\ work\end{tabular}}  & \multicolumn{4}{l|}{\cellcolor[HTML]{EFEFEF}Vortex + Extensions (CMF + 3xDMSLs + 3 dcache ports)}                                                                                                                                                                                                                                                                                                                                        \\ \hline \hline
\rowcolor[HTML]{FFFFFF} 
\textbf{Arch}                                                 & \multicolumn{1}{c|}{\cellcolor[HTML]{FFFFFF}\textbf{Cores}} & \multicolumn{1}{c|}{\cellcolor[HTML]{FFFFFF}\textbf{\begin{tabular}[c]{@{}c@{}}Area \\ (mm2)\end{tabular}}} & \multicolumn{1}{c|}{\cellcolor[HTML]{FFFFFF}\textbf{\begin{tabular}[c]{@{}c@{}}Perf. \\ (GFLOP/s)\end{tabular}}} & \multicolumn{1}{c|}{\cellcolor[HTML]{FFFFFF}\textbf{\begin{tabular}[c]{@{}c@{}}Area Eff. \\ (GFLOP/s/mm2)\end{tabular}}} \\ \hline
\rowcolor[HTML]{FFFFFF} 
VB                                                            & \multicolumn{1}{c|}{\cellcolor[HTML]{FFFFFF}1}              & \multicolumn{1}{r|}{\cellcolor[HTML]{FFFFFF}2.41}                                                           & \multicolumn{1}{r|}{\cellcolor[HTML]{FFFFFF}0.87}                                                                & 0.35                                                                                                                     \\ \hline
\rowcolor[HTML]{FFFFFF} 
VB                                                            & \multicolumn{1}{c|}{\cellcolor[HTML]{FFFFFF}2}              & \multicolumn{1}{r|}{\cellcolor[HTML]{FFFFFF}4.81}                                                           & \multicolumn{1}{r|}{\cellcolor[HTML]{FFFFFF}1.46}                                                                & 0.30                                                                                                                     \\ \hline
\rowcolor[HTML]{FFFFFF} 
VB                                                            & \multicolumn{1}{c|}{\cellcolor[HTML]{FFFFFF}4}              & \multicolumn{1}{r|}{\cellcolor[HTML]{FFFFFF}9.63}                                                           & \multicolumn{1}{r|}{\cellcolor[HTML]{FFFFFF}2.93}                                                                & 0.30                                                                                                                     \\ \hline
\rowcolor[HTML]{FFFFFF} 
VB                                                            & \multicolumn{1}{c|}{\cellcolor[HTML]{FFFFFF}8}              & \multicolumn{1}{r|}{\cellcolor[HTML]{FFFFFF}19.26}                                                          & \multicolumn{1}{r|}{\cellcolor[HTML]{FFFFFF}5.83}                                                                & 0.30                                                                                                                     \\ \hline

\rowcolor[HTML]{FFFFFF} 
$\mathrm{VU_4}$                                                           & \multicolumn{1}{c|}{\cellcolor[HTML]{FFFFFF}1}              & \multicolumn{1}{r|}{\cellcolor[HTML]{FFFFFF}2.41}                                                           & \multicolumn{1}{r|}{\cellcolor[HTML]{FFFFFF}0.91}                                                                & 0.38                                                                                                                     \\ \hline
\multicolumn{1}{|c|}{$\mathrm{VU_8}$}                                     & \multicolumn{1}{c|}{1}                                      & \multicolumn{1}{r|}{2.41}                                                                                   & \multicolumn{1}{r|}{0.90}                                                                                        & 0.37                                                                                                                     \\ \hline
\multicolumn{1}{|c|}{Spatz}                                     & \multicolumn{1}{c|}{2}                                      & \multicolumn{1}{r|}{0.74}                                                                                   & \multicolumn{1}{r|}{4.13}                                                                                        & 5.59                                                                                                                    \\ \hline
\multicolumn{1}{|c|}{Ara2}                                     & \multicolumn{1}{c|}{1}                                      & \multicolumn{1}{r|}{4.47}                                                                                   & \multicolumn{1}{r|}{3.88}                                                                                        & 0.87                                                                                                                     \\ \hline
\rowcolor[HTML]{EFEFEF} 
\textbf{\begin{tabular}[c]{@{}c@{}}This \\ work\end{tabular}} & \multicolumn{1}{c|}{\cellcolor[HTML]{EFEFEF}1}              & \multicolumn{1}{r|}{\cellcolor[HTML]{EFEFEF}2.89}                                                           & \multicolumn{1}{r|}{\cellcolor[HTML]{EFEFEF}\textbf{4.63}}                                                       & \textbf{1.63}                                                                                                            \\ \hline
\end{tabular}
}
    \vspace{-1em}
    \label{tab:comp}
\end{table}

%% file: text/7_conclusion.tex
\section{Conclusion}

In this work, we identify 
the control flow (CF) management and memory orchestration as the main performance overheads in the open-source Vortex GPGPU platform 
for regular memory-intensive kernels. To overcome this, we propose a set of microarchitectural enhancements of the Vortex GPGPU, 
allowing the GPGPU pipeline back-end utilization to improve from 10\% to 50\% on average. 
Furthermore, our evaluation shows an 8$\times$ faster execution, 10$\times$ dynamic instruction reduction, 
boosting the performance of the Vortex GPGPU from 0.35 to 1.63 $\mathrm{GFLOP/s/mm^2}$: 8 baseline cores would be needed to match 1 of our extensions-enhanced core.
Our proposed enhancements offer improvements, even when the number of cores, warps, and threads are scaled, thereby they can be adopted across a wide range of performance targets and area budgets. 
These enhancements can be integrated into application-level libraries in the future to unleash Vortex as a competitive GPGPU platform. 